%% file: arxiv_combined.tex
\documentclass[twocolumn,amsmath,amssymb,aps,prb,longbibliography,nofootinbib,superscriptaddress]{revtex4-2}

\usepackage{graphicx}% Include figure files
\usepackage{dcolumn}% Align table columns on decimal point
\usepackage{bm}% bold math
\usepackage{mathdots}
\graphicspath{ {./Figures/} }

\makeatletter
\let\frontmatter@footnotetext\@gobble
\makeatother

\newcommand{\dc}{\mathrm{dc}}
\newcommand{\crm}{\mathrm{c}}
\newcommand{\bb}{\mathrm{b}}
\newcommand{\g}{\mathrm{g}}
\newcommand{\N}{\mathrm{N}}
\newcommand{\D}{\mathrm{D}}
\newcommand{\m}{\mathrm{m}}
\newcommand{\s}{\mathrm{s}}

\begin{document}

\title{Aperiodic temporal modulation for distortionless broadband impedance matching beyond the Bode-Fano limit}

\author{Martin Olguin-Lopez}
\affiliation{Department of Electrical and Computer Engineering, University of Delaware, Newark, DE 19716, USA}

\author{Dimitrios L. Sounas}
\affiliation{Department of Electrical and Computer Engineering, Wayne State University, Detroit, MI 48202, USA}

\author{Nader Engheta}
\affiliation{Department of Electrical and Systems Engineering, University of Pennsylvania, Philadelphia, PA 19104, USA}

\author{Andrea Al\`u}
\affiliation{Photonics Initiative, Advanced Science Research Center, City University of New York, New York, NY 10031, USA}
\affiliation{Physics Program, Graduate Center, City University of New York, New York, NY 10016, USA}

\author{Mario J. Mencagli}
\altaffiliation{~}
\affiliation{Department of Electrical and Computer Engineering, University of Delaware, Newark, DE 19716, USA}

\date{\today}

\begin{abstract}
The Bode–Fano bound sets a fundamental trade-off between bandwidth and reflection in passive, linear, time-invariant matching networks. We show that an aperiodically time-modulated reactive element can emulate the non-Foster response required to match a prescribed pulse, achieving reflectionless, nearly distortionless energy transfer beyond the Bode–Fano limit. The approach introduces a new constraint: a minimum dc bias that scales with pulse bandwidth, derived from the requirement that the modulated capacitance remain positive at all times. This modulation-bias bound replaces the classical bandwidth-reflection trade-off with a bandwidth-energy trade-off. A realistic circuit simulation confirms broadband matching with preserved waveform fidelity, demonstrating that the scheme is physically realizable and not merely a mathematical circumvention.
\end{abstract}

\maketitle
The Bode–Fano (BF) bound \cite{bode,fano} establishes a fundamental limit on broadband impedance matching in electromagnetic engineering. It defines an inherent trade-off between bandwidth and matching efficiency: the bandwidth over which high-efficiency transfer of energy to a given load is maximized and reflections are minimized is fundamentally limited. This limitation is particularly significant for electrically small devices, as these tend to behave as highly reactive loads.

Importantly, two other fundamental electromagnetic limits---Chu's limit for small antennas \cite{Wheeler1947,Chu1948} and Rozanov's bound for thin absorbers \cite{Rozanov2000}---can ultimately be treated as impedance-matching problems, see, e.g., Refs.~\cite{Pfeiffer2017,manteghi2019,Firestein2023}. Therefore, the BF bound serves as an essential reference point for designing compact devices in fields such as medical equipment, portable wireless devices, communication systems, and sensors, where space constraints make broadband matching challenging.

To circumvent the BF bound, approaches that deliberately violate its underlying assumptions---linearity, passivity, and time invariance---and enable broadband matching have been actively pursued. Early efforts primarily focused on breaking the passivity assumption by using non-Foster (active) elements \cite{hrabar2011,zhu2012,Loghmannia2021}. However, non-Foster matching networks are well known to suffer from stability issues and to degrade the signal-to-noise ratio (SNR) \cite{munoz2012,jacob2016,shih2018}. More recently, renewed interest in time-varying media has opened the door to new wave phenomena, see, e.g., Refs.~\cite{Galiffi2022,Ortega2023,Ptitcyn2023,Koutserimpas2024,Lustig2018,Pacheco2020,Jayathurathnage2021,Mencagli2022,Wang2025}, and provided an alternative route to broadband matching by violating the time-invariance assumption. Time-modulation approaches can be divided into two categories, depending on whether they use periodic or non-periodic forms of modulation. Periodic modulation allows one to design matching networks that are independent of the input signals \cite{huanan2019,Fritts2025}. However, they need to access regimes of parametric gain, otherwise such networks follow bounds similar to the BF bound \cite{MTM_2022}. Parametric gain is more controllable than other active devices, but can still lead to instabilities. Aperiodic forms of temporal modulation can expand the return loss bandwidth beyond the BF bound without the problem of instabilities \cite{shlivinski2018,yang2022}, but they are not signal agnostic, i.e., they need to be timed with the specific  signal they need to impedance match. In addition, existing approaches are based on abrupt switching, and generally introduce distortion to the signal transmitted to the load.

\begin{figure}[t!]
    \centering
    \includegraphics[width=\columnwidth]{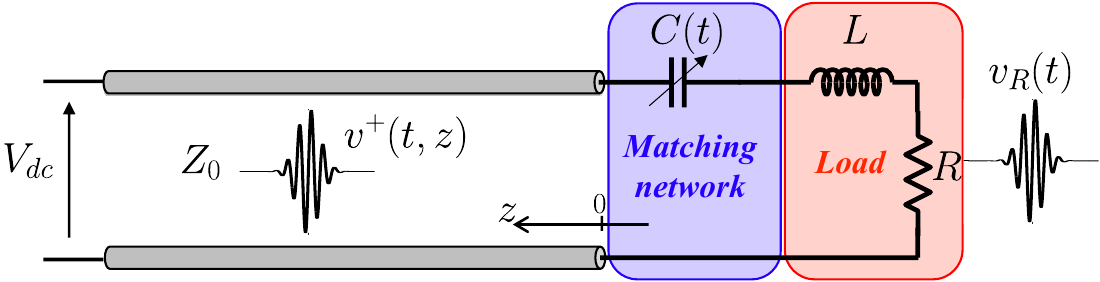}
    \caption{\label{matching_architecture}Matching of a reactive load (here modeled as a series combination of an inductor and a resistor) to a transmission line through an aperiodically-modulated capacitor.}
\end{figure}

In a different context, it has been shown that time modulation can be used to transform the response of a circuit for prescribed input signals \cite{hrabar2020,ptitcyn2023_2}, enabling also non-Foster responses. This approach also requires time synchronization between the modulation and the incoming signals. Here, we use aperiodic modulation to realize effective non-Foster responses for broadband matching of highly reactive loads. In addition to completely suppressing reflections, this approach minimizes the temporal distortion of the signal at the load, thereby preserving the waveform of the pulse to be transmitted or radiated.
% Meeting these combined requirements has remained challenging for prior approaches. 
Similarly to other approaches that transform the circuit response with time modulation \cite{hrabar2020,hecht2023,ptitcyn2023_2}, we need to bias with a nonzero static voltage the modulation waveform, in order to avoid nonphysical values in the modulated components. To understand the implications of this requirement, we derive bounds on the minimum required static bias versus the pulse bandwidth and energy for arbitrary pulse profiles. We validate the proposed matching approach in a realistic circuit implementation and transient nonlinear circuit simulations, showing that it can indeed surpass the BF bound.

\emph{Aperiodically Time-Modulated Matching Network}.--We begin by considering the canonical architecture modeling an impedance matching problem: a matching network between a transmission line and a load (Fig.~\ref{matching_architecture}). In the proposed scheme, the matching network is an aperiodically time modulated capacitor $C\left(t\right)$. The load is assumed to be inductive, a resistor $R$ in series with an inductor $L$, modeling, for example, an electrically small loop antenna. 
%It is well known that matching highly reactive loads over a wide bandwidth poses a major challenge \cite{bode,fano}. 
Transformers can equalize the real part of the load to the transmission-line characteristic impedance over a relatively broad bandwidth. Thus, we assume $Z_0=R$. The circuit is excited by a voltage pulse  $v^+(t)$ that propagates along the line towards the load, and the variable capacitor is connected to a DC source $V_{\dc}$.
%We further assume that an input voltage pulse [$v^+\left(t,z\right)$] starts to propagate along the transmission line toward the load after the capacitor in the matching network has been charged by a DC source ($V_{\dc}$), which remains connected to the system throughout. 
%A bias applied to a time-varying capacitor has also been used to explore other wave phenomena \cite{hrabar2020,hecht2023,ptitcyn2023_2,schwartz2025}.
Similarly to other approaches for the implementation of arbitrary circuit responses using time modulation \cite{hrabar2020,hecht2023,ptitcyn2023_2,schwartz2025}, a DC bias is necessary to avoid singularities or negative values in the time-varying capacitor.

By enforcing zero reflections and assuming that the voltage across the resistor is the same as the input pulse, i.e., $v_{R}(t)=v^+(t)$, Kirchhoff’s voltage law (KVL) yields $v_C(t) = V_{\dc} - v_L(t)$, with $v_{C}(t)$ and $v_{L}(t)$ denoting the voltage across the capacitor and the inductor, respectively. By combining the previous equation with the capacitor $q$-$v$ relation, we obtain the required temporal variation of the capacitance to fully transmit $v^+(t)$ to the resistor load without distortion \cite{supp}, 
\begin{equation}\label{Ct}
C\left(t\right) = 
\frac{
\frac{1}{R} \displaystyle\int_{-\infty}^{t} v^+(\nu)\, d\nu + V_{\dc} C_0
}{
\,V_{\dc} - \frac{L}{R}\, \frac{d v^+(t)}{dt}\,
}
\end{equation}
with $C_0$ denoting the initial value of the capacitance. Eq.~\eqref{Ct} can be written as $-L\frac{di^+(t)}{dt} = v_C(t) - V_{\dc}$, where $i_+(t)$ is the current flowing through the circuit and $v_C(t) - V_{\dc}$ is the alternating component of the capacitor voltage. This is the constitutive relation of a negative inductor $-L$ showing that the time-varying capacitor realizes such an inductance for the specified incident pulse. 

\begin{figure}
    \includegraphics[width=\columnwidth]{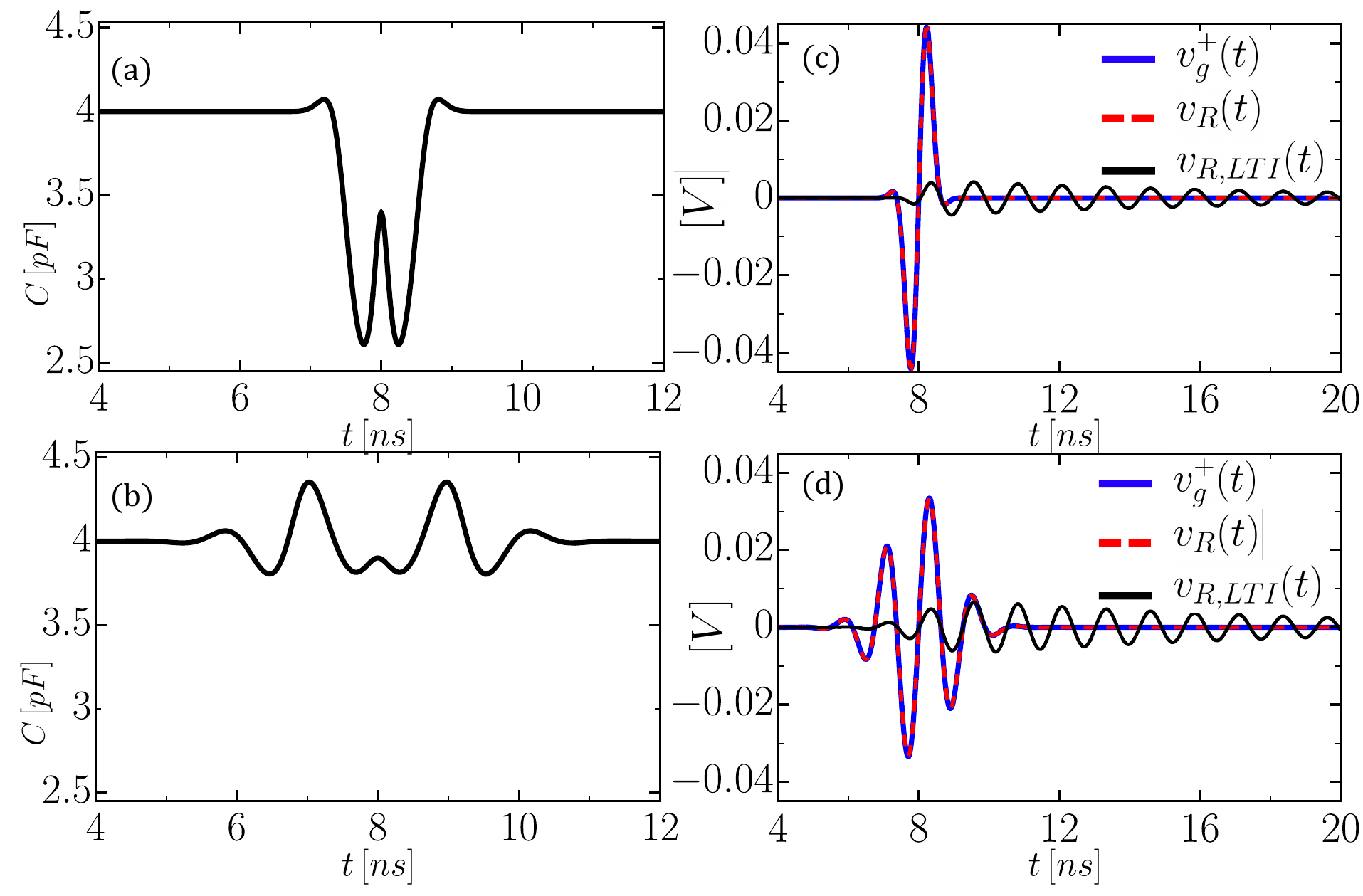}
    \caption{\label{figCt} Time-varying capacitance profiles and corresponding load responses for the matching architecture of Fig.~\ref{matching_architecture}. (a), (b) Temporal profiles of $C(t)$ for $\sigma = 0.3~\mathrm{ns}$ and $\sigma = 0.9~\mathrm{ns}$, respectively. (c), (d) Voltage across the load resistor $v_R(t)$. In each panel of (c) and (d), the input pulse $v_\g^{+}(t)$ and the load voltage for a linear time-invariant (LTI) network with $C(t)=C_0$ are also reported for comparison. The results were obtained for $C_0=4~\text{pF}$, $L=10~\text{nH}$, $R=1~\Omega$, $E_v=1~\text{pJ}$, $V_{\dc}=6~\text{V}$, and $\omega_\crm=\omega_0=1/\sqrt{C_0 L}$.} 
\end{figure}

As an illustrative example, we consider a modulated Gaussian pulse $v^{+}_{\g}(t)=A e^{-(t-\mu)^{2}/(2\sigma^{2})}\sin\!\left(\omega_{\crm} (t-\mu)\right)$, where $A$, $\mu$, $\sigma$, and $\omega_{\crm}$ denote the pulse amplitude, temporal center, temporal width, and carrier frequency, respectively. Replacing $v^+_{\g}(t)$ into Eq.~(\ref{Ct}), an analytical form for $C\left(t\right)$ can be obtained \cite{supp}. Figs.~\ref{figCt}(a) and (b) show the temporal profile of $C(t)$ for $\sigma = 0.3~\mathrm{ns}$ and $\sigma = 0.9~\mathrm{ns}$, respectively. As $\sigma$ increases, the bandwidth of $v_{\g}^{+}(t)$ decreases and, as expected, a smaller excursion of $C(t)$ is required to achieve matching. To validate the matching capabilities of these capacitance profiles, the architecture of Fig.~\ref{matching_architecture} is implemented in Keysight ADS \cite{Keysight_ADS}, with the time-varying capacitor realized through a linear symbolically defined device. Figs.~\ref{figCt}(c) and (d) show the voltage across the load resistor $v_R(t)$ for $\sigma = 0.3~\mathrm{ns}$ and $\sigma = 0.9~\mathrm{ns}$, respectively. In each plot, the input pulse $v_\g^{+}(t)$ and the load voltage $v_{R,\mathrm{LTI}}(t)$ for matching with a linear time-invariant (LTI) capacitor equal to $C_0$ are also reported for comparison. The parameters of the load are selected to give a characteristic decay time $\tau = L/R = 10$~ns, significantly larger than the characteristic time $\sigma$ of the pulse, thus representing a remarkably challenging matching scenario. In this case, the LTI configuration leads to strong reflections and a pronounced ringing response in the load voltage. In contrast, when the capacitor is modulated according to Eq.~(\ref{Ct}), the voltage across the load closely reproduces the input pulse, demonstrating distortionless matching.

The approach described above works for pulses of arbitrarily large bandwidths. However, as we will show next, a larger bandwidth requires the use of a larger $V_\dc$ to avoid nonphysical values for $C(t)$, such as negative values or infinities.
%We now turn to the conditions required to ensure the capacitance profile $C\left(t\right)$ remains physically realizable. 
To ensure that $C\left(t\right)$ remains finite and strictly positive for any $t$, $V_{\dc}$ must be chosen to prevent (i) singularities and (ii) sign changes in Eq.~(\ref{Ct}). These requirements yield two admissible branches for $V_{\dc}$, a positive and a negative branch, depending on whether the numerator and denominator in Eq.~(\ref{Ct}) are both positive or negative, respectively. The positive branch is defined by $V_{\dc} > V_{\dc}^{+} =\max\!\left\{V_{\dc}^{\N+},\, V_{\dc}^{\D+}\right\}$, where $V_{\dc}^{\N+}=-\frac{1}{R C_{0}}\operatorname*{inf}_{t}  \int_0^{t} v^{+}(\nu)\, d\nu$ and $V_{\dc}^{\D+}=\frac{L}{R}\operatorname*{sup}_{t}  \frac{d v^{+}(t)}{dt}$ are the lower limits on  $V_{\dc}$ for the numerator and denominator to be positive, respectively. 
%and $\operatorname*{sup}$ ($\operatorname*{inf}$) denotes the supremum (infimum). 
The negative branch is defined by $V_{\dc} < V_{\dc}^{-} = \min\!\left\{V_{\dc}^{\N-},\, V_{\dc}^{\D-}\right\}$, where $V_{\dc}^{\N-}=-\frac{1}{R C_{0}}\operatorname*{sup}_{t}  \int_0^{t} v^{+}(\nu)\, d\nu$ and $V_{\dc}^{\D-}=\frac{L}{R}\operatorname*{inf}_{t}  \frac{d v^{+}(t)}{dt}$ are the upper limits on  $V_{\dc}$ for the numerator and denominator to be negative, respectively. %As we show next, these conditions lead to trade-offs between the bandwidth of the incident pulse, i.e., the matching bandwidth of the system, and the bias of the variable capacitor.
\begin{figure}
    \includegraphics[width=\columnwidth]{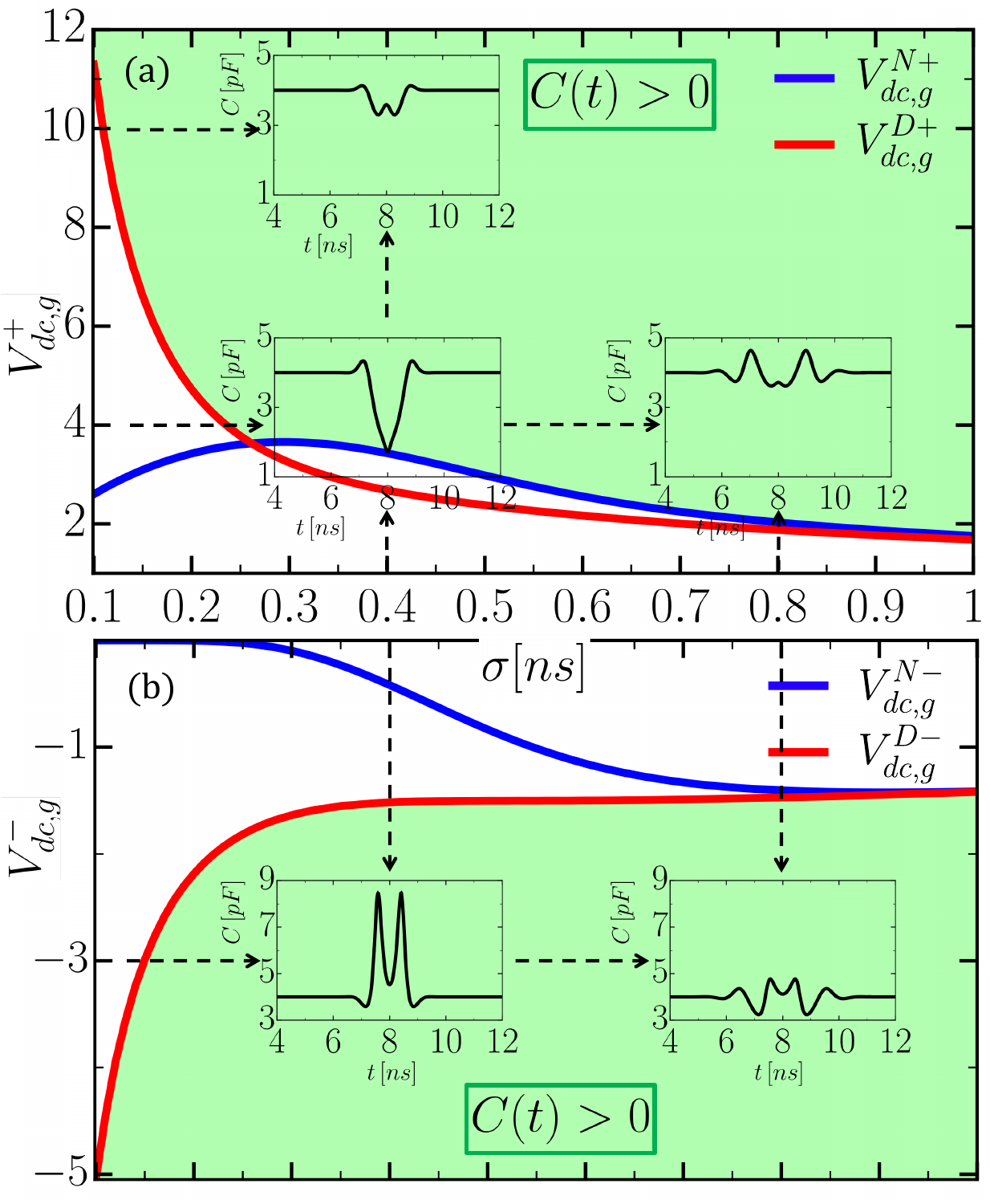}
    \caption{\label{Vdc_gauss} Admissible dc-bias voltage required to achieve distortionless impedance matching with $C\left(t\right)>0$ of a modulated Gaussian pulse as a function of its temporal width $\sigma$ (bandwidth $\propto 1/\sigma$). The green shaded region indicates the admissible set of voltage values for (a) the positive-bias branch and (b) the negative-bias branch. Insets show the profile of $C\left(t\right)$ for different $(\sigma,V_{\dc})$ pairs. The results were obtained for $C_0=4\text{pF}$, $L=10\text{nH}$, $R=1[\Omega]$, $E_v=1\text{pJ}$, and $f_c=f_0=\left(2 \pi \sqrt{C_0 L}\right)^{-1}$.} 
\end{figure}

To better understand the effect of the load and pulse characteristics on these limits, we derive $V_\dc^\pm$ for the modulated Gaussian pulse $v^{+}_{\g}(t)$ introduced earlier. For this pulse, $V_{\dc,\g}^{\N+}$ and $V_{\dc,\g}^{\D+}$ can be evaluated analytically \cite{supp} and are plotted versus versus $\sigma$ in Fig.~\ref{Vdc_gauss}(a). These results are derived for fixed load parameters and a fixed energy, $E_v$, for the pulse. The green regions in Fig.~\ref{Vdc_gauss} represent the allowed values of $V_{\dc}$. As $\sigma$ increases, $V_{\dc,\g}^{\N+}$ and $V_{\dc,\g}^{\D+}$ decrease, showing that the required $V_{\dc}$ decreases as the pulse bandwidth decreases. In the limiting case of large $\sigma$ (narrowband pulses), $V_{\dc,\g}^{\N+} \cong \frac{1}{\omega_{\crm} C_0 R} \sqrt{\frac{2 E_v}{\sigma \sqrt{\pi}}}$ and $V_{\dc,\g}^{\D+} \cong \frac{\omega_{\crm}L}{R} \sqrt{\frac{2 E_v}{\sigma \sqrt{\pi}}}$. This result shows that, for large $\sigma$, $V_\dc$ decreases as $\sigma^{-1/2}$. A similar behavior is observed for the negative branch [Fig.~\ref{Vdc_gauss}(b)]. In the opposite limiting case of small $\sigma$ (wide bandwidth),  $V_{\dc,\g}^{\N+}  \cong \sigma^{1/2} \frac{1}{C_0 R} \sqrt{\frac{4 E_v}{\sqrt{\pi}}}$ and $V_{\dc,\g}^{\D+}  \cong \sigma^{-3/2} \frac{L}{R} \sqrt{\frac{2 E_v}{\sqrt{\pi}}}$. The latter expression is strongly dominant, increasing as $\sigma$ decreases, while $V_{\dc,\g}^{\N+}$ exhibits the opposite trend [Fig.~\ref{Vdc_gauss}(a)]. A similar behavior is observed for the negative branch, with $V_{\dc,\g}^{\D-}$ being strongly dominant [Fig.~\ref{Vdc_gauss}(b)]. In either case, the minimum required $V_{\dc}$ scales as $\sigma^{-3/2}$. From Figs.~\ref{Vdc_gauss}(a) and (b), we also observe that, for a given $\sigma$, the negative branch $V_{\dc,\g}^{-}$ yields lower absolute DC bias voltages and is therefore preferable for designs targeting the minimum $V_\dc$. Finally, comparing the temporal profiles of $C\left(t\right)$ obtained from Eq.~(\ref{Ct}) for different $V_\dc$ and the same $v^{+}_{\g}(t)$ shows that selecting $V_{\dc}$ close to the limits leads to larger peak excursions for $C\left(t\right)$.

The bounds derived above show that $V_\dc^\pm\to 0$ for $\sigma\to\infty$. For pulses of fixed energy, this is a result of the fact that, in the limit $\sigma\to\infty$, the pulse amplitude goes to zero, essentially implying no input signal. A more realistic scenario in the limit $\sigma\to\infty$ is for pulses of constant amplitude $A$, which become continuous signals in the limit $\sigma\to\infty$. In such a case, it can be shown that $V_{\dc,\g}^\pm \cong c A$, with $c$ a constant that depends only on the load \cite{supp}. This seems to suggest that a nonzero $V_\dc$ is necessary even for continuous signals, contrary to what is expected from conventional matching theory, where a continuous signal can be perfectly matched to a reactive load by simply selecting $C_0$ to resonate with the load at the frequency of the signal. The resolution to this paradox comes from the fact that $V_\dc^\pm$ is derived under the very strict condition of zero reflection. As we show in \cite{supp}, in the limit $\sigma\to\infty$, $C(t) \to C_0$ and reflections stay low for a wide range of $V_\mathrm{dc}$ values that can be different from the $V_\mathrm{dc}$ used in the evaluation of $C(t)$ through Eq.~\eqref{Ct}. 

The previous analysis can be extended to arbitrary pulses by using Bernstein’s inequality \cite{lapidoth2009,pinsky2002}. To this end, we consider pulses with finite energy ($E_v<\infty$) and finite support $[\omega_1,\omega_2]$. Under these assumptions, Bernstein’s inequality combined with Fourier inversion and the Cauchy–Schwarz inequality \cite{lapidoth2009} allows us to derive a sufficient condition on the minimum biasing voltage as
\begin{subequations}\label{VdcB}
\begin{align}
|V_{\dc}|  &> \frac{1}{\omega_1 C_0 R} \sqrt{\frac{E_v B}{\pi}},  && \omega_0 \ge \omega_{\crm}, \label{VdcNB}\\
|V_{\dc}|  &> \frac{\omega_2 L}{R} \sqrt{\frac{E_v B}{\pi}}, && \omega_0< \omega_{\crm},  \label{VdcDB}
\end{align}
\end{subequations}
where $B = \omega_2 - \omega_1$ is the bandwidth and $\omega_{\crm} = \sqrt{\omega_1\omega_2}$ the center frequency of the pulse. The mathematical details of this derivation are provided in \cite{supp}. The bound predicts the same admissible regions for positive and negative $V_{\dc}$, which is expected from the waveform independence of the bound. 
%In contrast, the exact admissible regions for a given waveform are generally asymmetric, reflecting the specific temporal profile of the pulse, which is not captured by energy- and bandwidth-only bounds.
The admissible threshold exhibits the same scaling with respect to $E_v$ and $R$ as for the case of the modulated Gaussian pulse. In the limiting case of narrow bandwidth ($B \to 0$), $\omega_1 \cong \omega_2 \cong \omega_{\crm}$, and Eqs.~(\ref{VdcNB}) and (\ref{VdcDB}) reduce to $V_{\dc}  > \frac{1}{\omega_{\crm}RC_0} \sqrt{\frac{E_v B}{\pi}}$ for $\omega_0 \ge \omega_{\crm}$ and $V_{\dc}  > \frac{\omega_{\crm}L}{R} \sqrt{\frac{E_v B}{\pi}}$ for $\omega_0 < \omega_{\crm}$, respectively. Consistent with the behavior observed for the modulated Gaussian pulse, the bound monotonically decrease with decreasing bandwidth $B$. 
 
In the opposite limit of wide bandwidth, $\omega_1 \approx \omega_{\crm}^2/B$ and $\omega_2 \approx B$, and Eqs.~(\ref{VdcNB}) and (\ref{VdcDB}) reduce to $V_{\dc} > \frac{B^{3/2}}{\omega_{\crm}^2 C_0 R} \sqrt{\frac{E_v}{\pi}}$ for $\omega_0 \ge \omega_{\crm}$ and $V_{\dc} > \frac{B^{3/2}L}{R} \sqrt{\frac{E_v}{\pi}}$ for $\omega_0 < \omega_{\crm}$. This scaling behavior is the same as for the modulated Gaussian pulse (Fig.~\ref{Vdc_gauss}). This result shows that the proposed approach allows the matching of arbitrarily broadband pulses at the expense of a larger $V_\dc$.

\begin{figure*}[ht!]
    \centering
    \includegraphics[width=\textwidth]{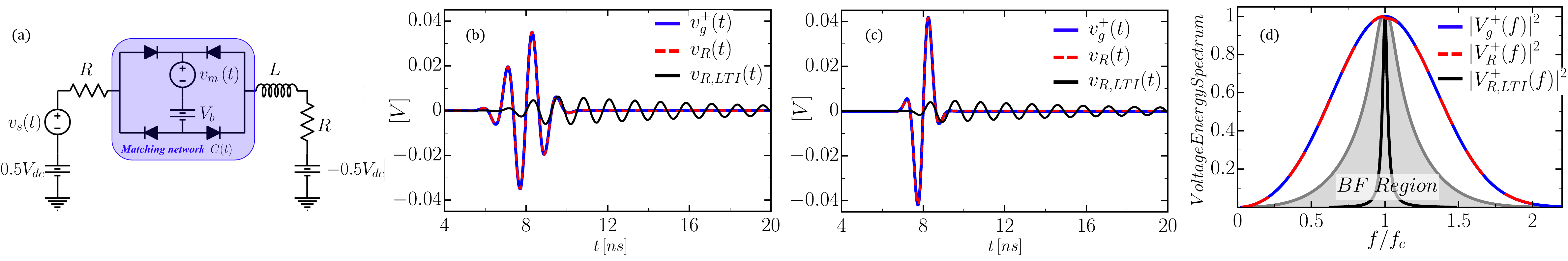}
    \caption{\label{Vr}Breaking the Bode-Fano bound with an aperiodically time-modulated matching scheme. (a) Proposed realistic implementation based on a differential Wheatstone bridge of four varactors. Temporal profile of the voltage across the load resistor obtained from a nonlinear transient simulation when the circuit in (a) is excited by a modulated Gaussian pulse with (b) $\sigma = 0.8\text{ns}$ and $V_{\dc}=2.1$, and (c) $\sigma = 0.4\text{ns}$ and $V_{\dc}=4$. In (b) and (c), the response of the corresponding linear time-invariant (LTI) case is also shown for comparison. (d) Output voltage energy spectrum for the input pulse in (c), with and without temporal modulation, including the Bode–Fano region. Other common parameters are the same as in Fig.~\ref{Vdc_gauss}.}
\end{figure*}

\emph{Modulation Energy}.--In lossless linear time-invariant (LTI) matching networks, energy transfer from the source to the load is maximized by minimizing the reflection coefficient. In contrast, in time-varying matching networks, there can be energy exchange between the modulation source and the system. As a result, even in the absence of reflection, the energy delivered to the load may differ from the incident energy, as part of it can be exchanged with the modulation source. 
For the circuit analyzed here, the net modulation energy over the entire process is $W_\mathrm{mod}=\frac{1}{2}V_{\dc} \left(q_\infty-q_0\right)$, where $q_0$ and $q_\infty$ are the capacitor charges before and after the transmission of the incident pulse to the load, respectively. This expression for $W_\mathrm{mod}$ is a direct consequence of energy conservation and, therefore, does not depend on the shape of the incoming pulse. Instead, it depends only on the difference of the capacitor energy before and after the pulse, which is given by one half the product of $V_\dc$ with $\Delta q$. By charge conservation, this variation is related to the current of the incoming pulse as $\Delta q=\int_{-\infty}^{\infty} i^{+}(t)\, dt$. Thus, when the incoming pulse carries zero net charge ($\Delta q= 0$), the modulation energy vanishes. This is the case of the modulated Gaussian pulse considered earlier. As shown in \cite{supp}, during the transient, the capacitor charge increases and decreases by equal amounts relative to its initial value $q_0$, resulting in $\Delta q= 0$. The corresponding temporal evolution of the modulation energy indicates that the modulation source injects and extracts the same amount of energy from the system \cite{supp}, yielding $W_\mathrm{mod}=0$. In contrast, if the incoming pulse carries a nonzero net charge, such as a Gaussian pulse, $\Delta q \ne 0$, leading to a nonzero modulation energy. This energy exchange between the system and the modulation source arises from the interplay between the DC bias and the incoming pulse. Specifically, a positive (negative) net flow of charge leads to an increase (decrease) of stored charge in the capacitor, and to energy extraction (injection) by the modulation source \cite{supp}.

\emph{Proposed Practical Implementation}.--To validate the main features of the proposed time-varying matching scheme, we consider the realistic implementation in Fig.~\ref{Vr}(a), consisting of a differential circuit in which the time-varying capacitor is realized by using a Wheatstone bridge of four varactors. The initial equivalent capacitance is set by the bias voltage $V_\bb$ and subsequently modulated in time through the control signal $v_\m(t)$. The input signal $v_\s(t)$ is equal to the modulated gaussian pulse $v_\g(t)$ introduced earlier and models the forward-propagating pulse $v^{+}(t)$ in the transmission line model of Fig.~\ref{matching_architecture}. The circuit is analyzed by using a commercial transient circuit simulator. The nonlinear model of the varactor (inverse square root model of an inversely biased diode) is used to determine the modulation signal $v_\m(t)$ that produces the required capacitance profile $C(t)$. Figs.~\ref{Vr}(b) and (c) show the voltage across the load resistor, $v_{R}(t)$, obtained from the nonlinear transient circuit simulations for two different values of $\sigma$. An excellent agreement with the input pulse $v^{+}_{\g}(t)$ can be observed, demonstrating the capability of the proposed approach to achieve matching without distorting the signal. 

On the other hand, matching with a time-invariant capacitor leads to a long damping time, strong reflections and a long ringing response in the voltage across the resistor $v_{R,\mathrm{LTI}}(t)$ [continuous solid black line in Figs.~\ref{Vr}(b) and (c)]. 

Further insight into the spectral characteristics of the output signal and the ability of the proposed approach to surpass the BF bound can be gained by examining the energy spectrum of the output voltage. In Fig.~\ref{Vr}(d), the gray-shaded region denoted as the BF region represents the area where an output pulse is bound to reside for any low-ripple passive time-invariant matching network as predicted from the Bode–Fano bound. The input voltage spectrum $|V_{\g}^{+}(f)|^{2}$ is also shown, together with the output spectrum $|V_{R,\mathrm{LTI}}(f)|^{2}$ of matching with a time-invariant capacitor $C_0$ and the output spectrum $|V_{R}(f)|^{2}$ for the proposed time-modulated matching network. The unmodulated response, as expected, lies within the BF region, whereas the time-modulated response extends to a large degree beyond the bound, demonstrating a significant enhancement of the matching bandwidth beyond the BF bound.

\emph{Conclusion}.--In this Letter, we introduced a time-modulated matching framework that enables distortionless broadband impedance matching beyond the limitations imposed by the BF bound. By employing aperiodic temporal modulation of reactive elements, the proposed approach enables broadband, reflectionless energy transfer while preserving the temporal profile of the incident pulse. Our analysis establishes a lower bound on an electric dc biasing required to ensure physically realizable modulation and provides an in-depth formulation for the modulation energy that applies to a large variety of pulses. We validated our results through a realistic circuit implementation, showing that the resulting output spectra extend beyond the BF bound. The approach requires prior knowledge of the incident waveform and synchronization between the waveform and the prescribed temporal modulation; however, the analysis in \cite{supp} shows that the normalized reflected energy is first-order insensitive to small synchronization errors. Because the RF and modulation-control paths are, in principle, independently generated, synchronization need not rely on feedback and may instead be established using a common timing reference. This differs from many non-Foster implementations, in which the synthesized impedance is realized through an active feedback network and is therefore directly subject to loop-stability constraints. These findings open new opportunities for broadband impedance matching in a wide range of electromagnetic systems, including antennas, RF front ends, and photonic platforms. More broadly, this framework offers a promising pathway to revisit and potentially overcome classical performance limits in electromagnetic systems.

\emph{Acknowledgments}.-- N.E. and A.A. acknowledge partial support from the Simons Foundation/Collaboration on Symmetry-Driven Extreme Wave Phenomena (Grant No. SFI-MPS-EWP-00008530-04). A.A. and D.L.S. acknowledge support from the Air Force Office of Scientific Research.

\par\bigskip
{\small
\noindent$^{*}$Corresponding author.\par
\noindent\texttt{mencagli@udel.edu}\par
}
\par\medskip
\bibliography{mybib}

% ==========================
% Supplemental Material
% ==========================

\clearpage
\onecolumngrid

\begin{center}
{\large\bfseries Supplemental Material}\\[6pt]
{\large Aperiodic Temporal Modulation for Distortionless Broadband
Impedance Matching Beyond the Bode--Fano Limit}
\end{center}

\bigskip

% Supplemental numbering
\renewcommand{\thesection}{S\arabic{section}}
\renewcommand{\thesubsection}{S\arabic{section}.\arabic{subsection}}
\renewcommand{\thefigure}{S\arabic{figure}}
\renewcommand{\thetable}{S\arabic{table}}
\renewcommand{\theequation}{S\arabic{equation}}

\setcounter{section}{0}
\setcounter{figure}{0}
\setcounter{table}{0}
\setcounter{equation}{0}

\input{supplement_body}

\providecommand{\noopsort}[1]{}\providecommand{\singleletter}[1]{#1}%

\end{document}

%% file: supplement_body.tex
\section{Derivation of the Capacitance Profile for Distortionless Matching}
\begin{figure}[ht]
  \centering
  \includegraphics[width=0.5\linewidth]{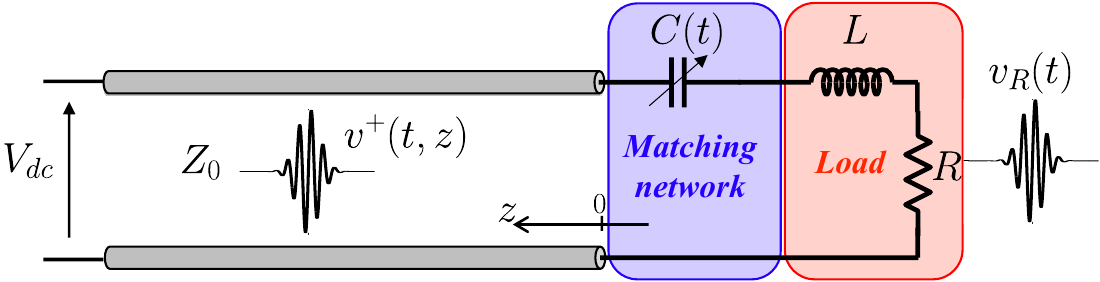}
  \caption{Matching of a reactive load (here, a series combination of an inductor and a resistor) to a transmission line through an aperiodically-modulated capacitor.}
  \label{figS1}
\end{figure}

Consider the aperiodically time-modulated impedance-matching architecture proposed in this manuscript, shown in Fig.~1 of the main text and reproduced here as Fig.~\ref{figS1} for convenience. The matching network, consisting of a capacitor actively modulated in time [$C(t)$], connects an $RL$ load to a feeding transmission line, which carries a voltage pulse $v^+\left(t,z\right)$ and biases the capacitor through a DC source $V_{\dc}$. To focus on reactive matching, which represents the main challenge \cite{bodeSM,fanoSM}, we assume $Z_0=R$, where $Z_0$ is the characteristic impedance of the transmission line. Applying the Kirchhoff’s voltage law (KVL) at the load, we obtain
\begin{equation}\label{KVL}
V_{\dc}+v^+(t)+v^-(t)=v_C(t)+v_L(t)+v_R(t),
\end{equation}
where $v^+(t)=v^+(t,z=0)$ and $v^-(t)=v^-(t,z=0)$ denote the incident and reflected pulses, respectively, and $v_C(t)$, $v_L(t)$, and $v_R(t)$ denote the voltage across the capacitor, inductor, and resistor, respectively. 
Imposing zero reflection, $v^-(t)=0$, and assuming that the voltage across the resistor is equal to the input pulse, $v_{R}\left(t\right)=v^+(t)$, the previous equation reduces to 
\begin{equation}\label{vc}
v_C(t)=V_{\dc}-v_L(t).
\end{equation}
where $v_L(t)=Ldi^+(t)/dt$ and $i^+(t)=v^+(t)/R$ is the current flowing into the load. The capacitance $C(t)$ can be determined from $C(t) = q(t)/v_C(t)$, where $q(t)$ is the capacitor charge given by
\begin{equation}\label{eq:q}
    q(t) = \int_{-\infty}^{t} i_C(\nu)\, d\nu + V_{\dc} C_0,
\end{equation}
$i_C(t)$ is the capacitor current, and $V_{\dc} C_0$ is the initial capacitor charge. Taking into account that in the absence of reflection, $i_C(t) = i^+(t) = v^+(t)/R$, we find
\begin{equation}\label{CtSM}
C\left(t\right) = 
\frac{
\frac{1}{R} \displaystyle\int_{-\infty}^{t} v^+(\nu)\, d\nu + V_{\dc} C_0
}{
\,V_{\dc} - \frac{L}{R}\, \frac{d v^+(t)}{dt}\,
}
\end{equation}

The previous equation enables an interesting parallel to be drawn between the proposed aperiodically time-modulated matching approach and the standard linear time-invariant (LTI) impedance-matching concept. In the LTI regime, the load of Fig.~\ref{figS1} is matched to the feeding transmission line with $Z_0=R$ when the capacitive impedance of the matching network cancels the inductive impedance of the load. As is well known, this cancellation occurs only at a single frequency, thereby inherently limiting the achievable matching bandwidth. In contrast, in the proposed time-modulated scheme, this condition is enforced in the time domain through the dynamic adjustment of the capacitance $C(t)$, which effectively makes the capacitor behave as a negative inductor, as can be seen from Eq.~(\ref{vc}), enabling broadband matching while preserving the temporal profile of the incoming pulse. 
%To determine $C(t)$, $v_C(t)$, given by Eq.~(\ref{vc}), and $i^+(t)=v^+(t)/R$ are substituted into the capacitor current-voltage relation, resulting in a first-order differential equation for $C(t)$, which is solved for $C(t)$ with the initial condition $C(0)=C_0$, yielding

From Eq.~\eqref{CtSM}, as discussed in the main text, two admissible branches for $V_{\dc}$, corresponding to a positive or negative $V_{\dc}$, can be defined to ensure $C(t)$ remains finite and strictly positive for any $t$. For convenience, we explicitly state these branches here. The positive branch is given by $V_{\dc}^{+} >\max\!\left\{V_{\dc}^{\N+},\, V_{\dc}^{\D+}\right\}$, where 
\begin{equation}\label{VdcPN}
V_{\dc}^{\N+}  =  -\frac{1}{R C_{0}}\operatorname*{inf}_{t\in \mathbb{R}}  \int_{-\infty}^{t} v^{+}(\nu)\, d\nu,
\end{equation}
and
\begin{equation}\label{VdcPD}
V_{\dc}^{\D+}  = \frac{L}{R}\operatorname*{sup}_{t\in \mathbb{R}}  \frac{d v^{+}(t)}{dt}.
\end{equation}
The negative branch is given by $V_{\dc}^{-} < \min\!\left\{V_{\dc}^{\N-},\, V_{\dc}^{\D-}\right\}$, where
\begin{equation}\label{VdcNN}
V_{\dc}^{\N-}  =  -\frac{1}{R C_{0}}\operatorname*{sup}_{t\in \mathbb{R}}  \int_{-\infty}^{t} v^{+}(\nu)\, d\nu,
\end{equation}
and
\begin{equation}\label{VdcDN}
V_{\dc}^{\D-}  = \frac{L}{R}\operatorname*{inf}_{t\in \mathbb{R}}  \frac{d v^{+}(t)}{dt}.
\end{equation}
$V_{\dc}^{+}$ and $V_{\dc}^{-}$ represent the lower and upper bounds, respectively, of the biasing level required to ensure that $C(t)$ remains finite and strictly positive.
%%%%%%%%%%%%%%%%%%%
% Modulated Gaussian Pulse
%%%%%%%%%%%%%%%%%%%
\section{Analysis for a Modulated Gaussian Input Pulse}
The input voltage is a modulated Gaussian pulse, as defined in the main text, given by
\begin{equation}\label{VGauss}
v^{+}_{\g}(t)=A e^{-(t-\mu)^{2}/(2\sigma^{2})}\sin\!\left(\omega_c (t-\mu)\right),
\end{equation}
where $A$, $\mu$, $\sigma$, and $\omega_c=2\pi\!f_c$ denote the pulse amplitude, temporal center, temporal width, and carrier angular frequency, respectively. Substituting $v^{+}_{\g}(t)$ in Eq.~(\ref{CtSM}), we obtain the required temporal capacitance profile $C(t)$, given by
\begin{equation}\label{CtGauss}
C\left(t\right)=\frac{R V_{\dc} C_0-
A e^{-\omega_c^{2}\sigma^{2}/2} \sqrt{\frac{\pi}{2}}\, \sigma
\operatorname{Im}\!\left\{
\operatorname{Erf}\!\left(\frac{t-\mu + i \omega_c \sigma^{2}}{\sqrt{2}\sigma}\right)
\right\}}
{R V_{\dc}
-
A L e^{-\frac{(t-\mu)^2}{2\sigma^2}}
\left[
\omega_c \cos\!\left(\omega_c (t-\mu)\right)
-
\frac{t-\mu}{\sigma^2}
\sin\!\left(\omega_c (t-\mu)\right)
\right]
}
\end{equation}
where $\operatorname{Erf}(x)=\frac{2}{\sqrt{\pi}}\int_{0}^{x} e^{-u^{2}}\,du$ is the error function, $\operatorname{Im}$ denotes the imaginary part of its argument, and $i=\sqrt{-1}$ denotes the imaginary unit. Having established the expression of $C(t)$ for the modulated Gaussian pulse, we now examine the admissible biasing values that ensure $C(t)$ remains finite and strictly positive, under different normalization conditions.
% Constant amplitude
\subsection{Admissible Biasing Values Under Constant-Amplitude Normalization}
Under this normalization, $\sigma$ is varied while the pulse amplitude $A$ is kept fixed, and the corresponding admissible biasing values are determined. 

It can be shown that $V_{\dc,\g}^{\N+}$ and $V_{\dc,\g}^{\D+} $ [Eqs.~(\ref{VdcPN}) and (\ref{VdcPD})], which define the positive lower bound $V_{\dc,\g}^{+}$ for the pulse $v^{+}_{\g}(t)$, are obtained at $t=\mu$ and are given by
\begin{equation}\label{VdcNPGauss}
V_{\dc,\g}^{\N+}  =  A\sqrt{\frac{\pi}{2}}\frac{\sigma}{R C_{0}}e^{-\frac{\omega_c^{2} \sigma^{2}}{2}}\operatorname{Erfi}\left(\frac{\omega_c\sigma}{ \sqrt{2}}\right)
\end{equation}
and
\begin{equation}\label{VdcDPGauss}
V_{\dc,\g}^{\D+}  = A \omega_c \frac{L}{R},
\end{equation}
where $\operatorname{Erfi}(x)=-i\operatorname{Erf}(ix)$ denotes the imaginary error function.
\begin{figure}[t]
  \centering
  \includegraphics[width=0.9\linewidth]{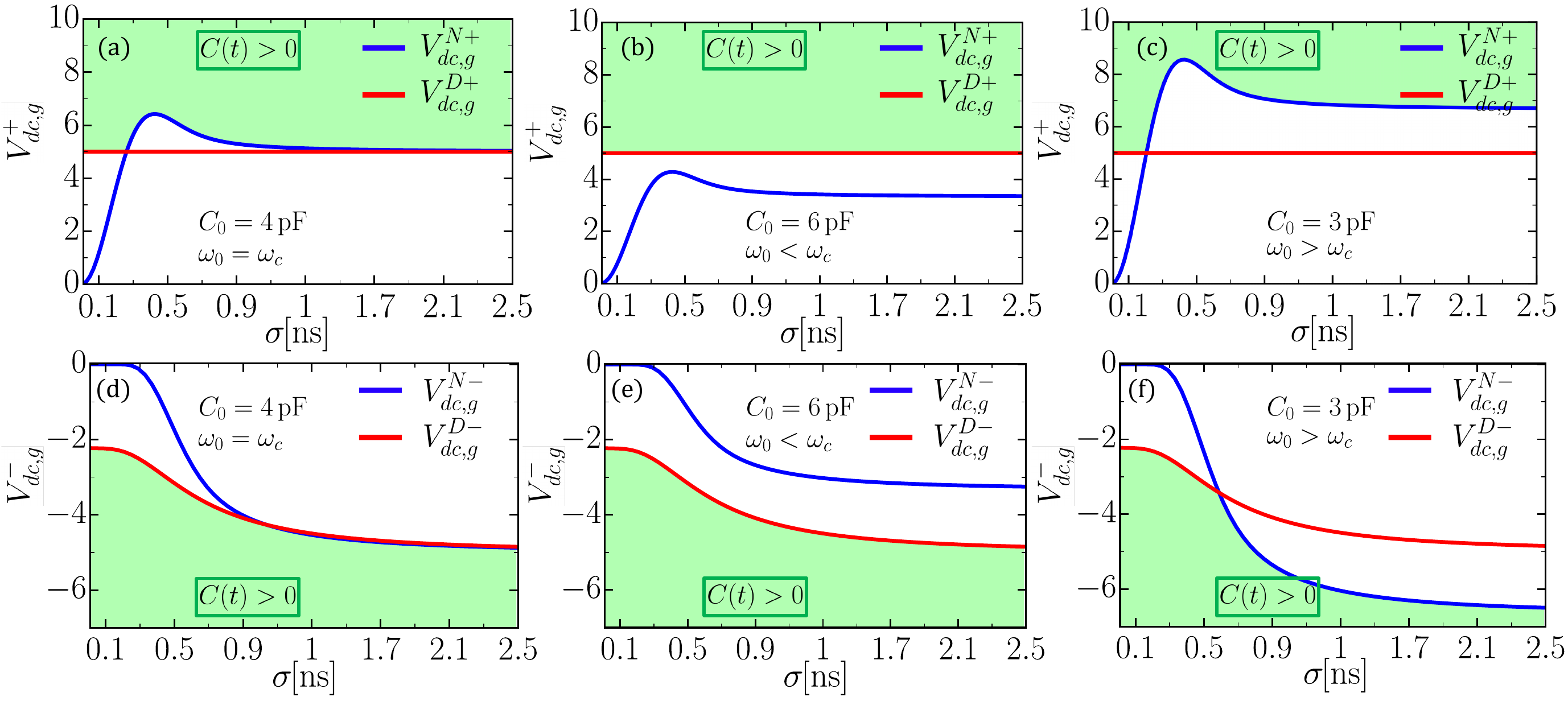}
  \caption{Admissible biasing values (constant-amplitude normalization). Panels (a)-(c) show the positive branch, $V_{\dc,g}^{+} >\max\!\left\{V_{\dc,g}^{N+},\, V_{\dc,g}^{D+}\right\}$, while panels (d)-(f) show the negative branch $V_{\dc,g}^{-} <\min\!\left\{V_{\dc,g}^{N-},\, V_{\dc,g}^{D-}\right\}$. The results were obtained for $A=0.1$[V], $L=10$[nH], $R=1[\Omega]$, $\omega_c=5$rad/ns, and $\omega_0=\left(\sqrt{C_0 L}\right)^{-1}$.}
  \label{figS2}
\end{figure}
Figs.~\ref{figS2}(a)-(c) show $V_{\dc,\g}^{\N+} $ and $V_{\dc,\g}^{\D+}$ as functions of $\sigma$, obtained from Eqs.~(\ref{VdcNPGauss}) and (\ref{VdcDPGauss}), respectively, for three different values of $C_0$. We observe that $V_{\dc,\g}^{\D+}$ is independent of both $C_0$ and $\sigma$, and therefore assumes the same value in Figs.~\ref{figS2}(a)–(c) (solid red curves). This behavior follows from the fact that $V_{\dc,\g}^{D+}$ is determined by the derivative of the incoming pulse. In particular, the supremum of the derivative of $v^{+}_{\g}(t)$ [Eq.~(\ref{VdcPD})] occurs at the center of the Gaussian envelope ($t=\mu$), where $v^{+}_{\g}(t)$ is dominated by the carrier frequency. As a result, $\sigma$ controls only the temporal extent of the Gaussian envelope, not its peak instantaneous slope.

On the other hand, the condition defining $V_{\dc,\g}^{\N+}$ [Eq.~(\ref{VdcPN})], which is inversely proportional to $C_0$, involves the integral of $v^{+}_{\g}(t)$ and depends on the interplay between the width of the Gaussian envelope and the carrier oscillations. As a result, $V_{\dc,\g}^{\N+}$, which also occurs at $t=\mu$, captures both the pulse width and the oscillatory regime. As shown in Figs.~\ref{figS2}(a)–(c) (solid blue curves), varying $C_0$ shifts the range of values of $V_{\dc,\g}^{\N+}$, whereas the variation within that range depends on $\sigma$. In the limits $\sigma \to 0$ and $\sigma \to \infty$, Eq.~(\ref{VdcNPGauss}) reduces to 
\begin{subequations}\label{VdcNPGaussAsym}
\begin{align}
V_{\dc,\g}^{\N+}   &= A\frac{\omega_c}{C_0 R}\sigma^2,  && \sigma \to 0, \label{VdcNPGaussAsymWide}\\
V_{\dc,\g}^{\N+}   &= \frac{A}{\omega_c C_0 R}, && \sigma \to \infty.  \label{VdcNPGaussAsymNar}
\end{align}
\end{subequations} 
showing that $V_{\dc,g}^{\N+}$ vanishes quadratically with $\sigma$ in the former case and becomes independent of $\sigma$ in the latter.

Combining Eqs.~(\ref{VdcNPGauss}) and (\ref{VdcDPGauss}) with the condition  $V_{\dc,\g}^{+} >\max\!\left\{V_{\dc,\g}^{\N+},\, V_{\dc,\g}^{\D+}\right\}$, we obtain the admissible sets of biasing values, corresponding to the green shaded regions in Figs.~\ref{figS2}(a)--(c). As $\sigma \to 0$, corresponding to an ultra-wideband pulse $v^{+}_{\g}(t)$ with bandwidth $B\sim1/\sigma$, the minimum required biasing level is determined by $V_{\dc,\g}^{\D+}$, since $V_{\dc,\g}^{\N+}$ approaches zero [Eq.~(\ref{VdcNPGaussAsymWide})]. It is interesting that in this case, $V_{\dc}^+$ depends only on the load, and increasing the pulse bandwidth does not lead to an increase in $V_{\dc}^+$. On the other hand, as $\sigma $ increases beyond the broadband regime, including the intermediate and narrowband cases, the minimum biasing depends on $C_0$ through the relation between the resonance angular frequency, $\omega_0=1/\sqrt{C_0 L}$, and the carrier angular frequency, $\omega_c$. For the value of $C_0$ such that $\omega_0=\omega_c$, $V_{\dc,\g}^{+}$, away from the broadband regime, is dominated by $V_{\dc,\g}^{\N+}$, which converges to $V_{\dc,\g}^{\D+}$ in the narrowband regime. For $C_0$ values such that $\omega_0<\omega_c$, $V_{\dc,\g}^{+}$ is determined by $V_{\dc,\g}^{\D+}$ for all values of $\sigma$. Conversely, for $C_0$ such that $\omega_0>\omega_c$, $V_{\dc,\g}^{+}$ is by determined by $V_{\dc,\g}^{\N+}$, except in the broadband regime.

Figs.~\ref{figS2}(d)--(f) show, for the same three values of $C_0$ as in Figs.~\ref{figS2}(a)--(c), the admissible set of biasing values for the negative branch, given by $V_{\dc,\g}^{-} <\min\!\left\{V_{\dc,\g}^{\N-},\, V_{\dc,\g}^{\D-}\right\}$, where $V_{\dc,\g}^{\N-}$ and $V_{\dc,\g}^{\D-}$ are obtained numerically from Eqs.~(\ref{VdcNN}) and (\ref{VdcDN}), respectively. In this branch, $V_{\dc,\g}^{\D-}$ depends on $\sigma$, since, in contrast to the supremum [Eq.~(\ref{VdcPD})], the infimum of the derivative of $v^{+}_{\g}(t)$ [Eqs.~(\ref{VdcDN})] occurs at a time instant that depends on $\sigma$. Apart from this difference, considerations analogous to those for the positve branch apply.
% Constant energy
\subsection{Admissible Biasing Values Under Constant-Energy Normalization}
\begin{figure}[ht]
  \centering
  \includegraphics[width=0.9\linewidth]{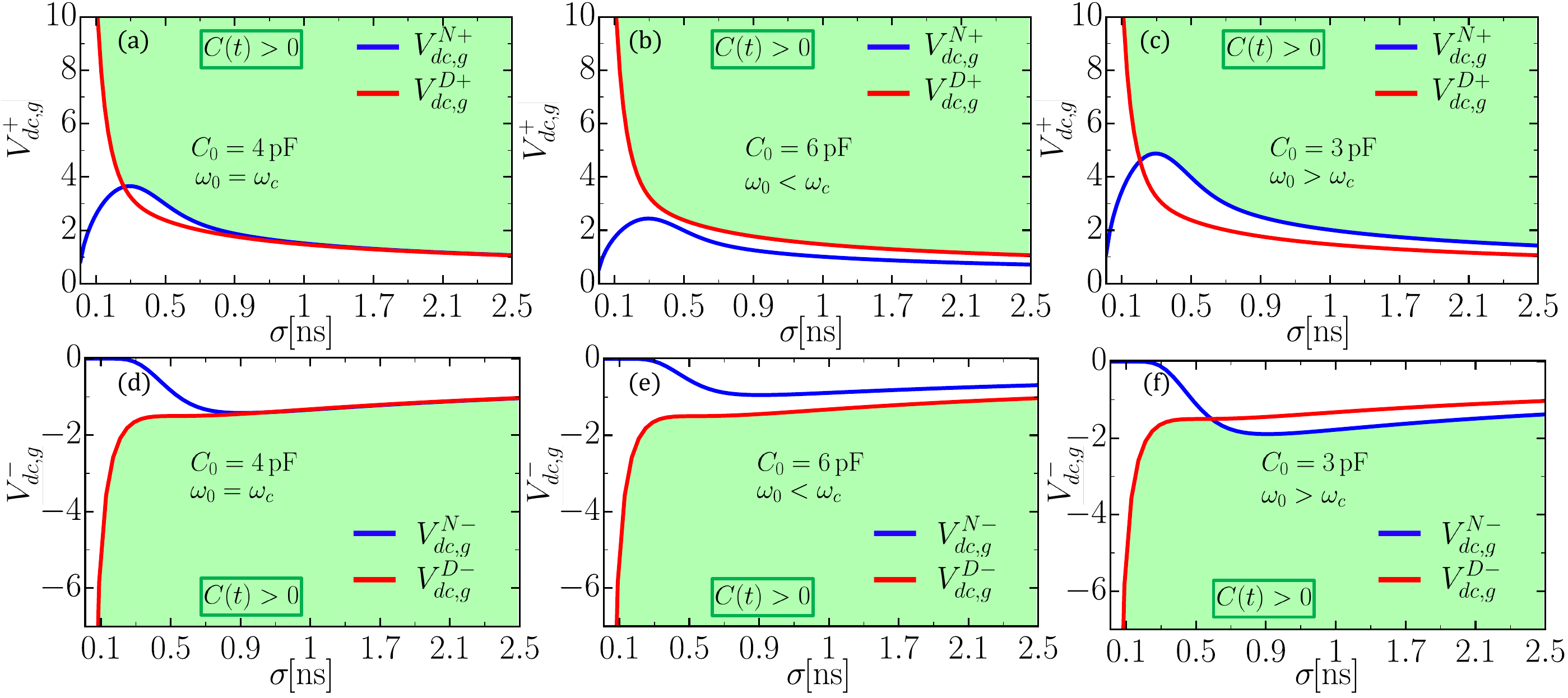}
  \caption{Admissible biasing values (constant-energy normalization). Panels (a)--(c) show the positive branch, $V_{\dc,\g}^{+} >\max\!\left\{V_{\dc,\g}^{\N+},\, V_{\dc,\g}^{\D+}\right\}$, while panels (d)--(f) show the negative branch $V_{\dc,\g}^{-} <\min\!\left\{V_{\dc,\g}^{\N-},\, V_{\dc,\g}^{\D-}\right\}$. The results were obtained for $E_v=1$ [pJ], $L=10$ [nH], $R=1~[\Omega]$, $\omega_c=5$ rad/ns, and $\omega_0=1/\sqrt{C_0 L}$.}
  \label{figS3}
\end{figure}

The results in Fig.~\ref{figS2} are obtained under constant-amplitude normalization, in which the pulse amplitude is kept fixed while varying $\sigma$, and the energy therefore changes accordingly. To decouple the effects of bandwidth and energy, we next consider the case in which the input pulse is normalized with respect to its energy, $E_v$. In this case, $\sigma$ is varied while the pulse amplitude is adjusted to maintain constant energy. To this end, the amplitude $A$ of $v^{+}_{g}(t)$ is expressed as
\begin{equation}\label{AnormE}
A = \sqrt{\frac{2E_v}{\sigma \sqrt{\pi}} \frac{1}{1 - e^{-\omega_c^2 \sigma^2}}}
\end{equation}
to ensure constant energy. Using this normalization and the same procedure as in Fig.~\ref{figS2} for the same values of $C_0$, we reanalyze the admissible biasing values for both branches. In particular, by substituting Eq.~(\ref{AnormE}) into Eqs.~(\ref{VdcNPGauss}) and (\ref{VdcDPGauss}), we obtain: %the expressions of $V_{\dc,g}^{N+}$ and $V_{\dc,g}^{D+}$:
\begin{subequations}\label{VdcGaussNorm}
\begin{align}
V_{\dc,\g}^{\N+}  &= \frac{1}{R C_{0}} \sqrt{\frac{\pi \sigma}{2 - 2e^{-\omega_c^2 \sigma^2}}}e^{-\frac{(\omega_c \sigma)^{2}}{2}}\operatorname{Erfi}\left(\frac{\omega_c\sigma}{ \sqrt{2}}\right)\sqrt{\frac{2E_v}{\sqrt{\pi}}}, \label{VdcGaussNNorm}\\
V_{\dc,\g}^{\D+}  &= \omega_c \frac{L}{R} \sqrt{\frac{1}{\sigma - \sigma e^{-\omega_c^2 \sigma^2}}} \sqrt{\frac{2E_v}{\sqrt{\pi}} }.\label{VdcGaussDNorm}
\end{align}
\end{subequations}
which are then combined into $V_{\dc,\g}^{+} >\max\!\left\{V_{\dc,\g}^{\N+},\, V_{\dc,\g}^{\D+}\right\}$ to determine the admissible set of positive biasing values. The admissible negative biasing values, $V_{\dc,\g}^{-} <\min\!\left\{V_{\dc,\g}^{\N-},\, V_{\dc,\g}^{\D-}\right\}$, are obtained by evaluating numerically Eqs.~(\ref{VdcNN}) and (\ref{VdcDN}). The green-shaded regions in Fig.~\ref{figS3} show the corresponding admissible values under constant-energy normalization. Since considerations analogous to those for the constant-amplitude normalization apply, particularly regarding the dependence on $C_0$ and outside the broadband and narrowband regimes, we focus on these two cases. In the broadband regime, while $V_{\dc,\g}^{\N+}$ tends to zero as $\sigma\to 0$, similar to the constant-amplitude case, $V_{\dc,\g}^{\D+}$ diverges [Fig.~\ref{figS3}(a)-(c)]. Specifically, under the assumption $\sigma \to 0$, Eqs.~(\ref{VdcGaussNNorm}) and (\ref{VdcGaussDNorm}) reduce to
\begin{subequations}\label{VdcGaussNorm}
\begin{align}
V_{\dc,\g}^{\N+}  &= \frac{\sqrt{2 \sigma}}{R C_{0}}\sqrt{\frac{2E_v}{\sqrt{\pi}}}, \label{}\\
V_{\dc,\g}^{\D+}  &= \frac{\sqrt{2}L}{\sigma^{3/2} R} \sqrt{\frac{2E_v}{\sqrt{\pi}} }, \label{}
\end{align}
\end{subequations}
from which it follows that $V_{\dc,\g}^{\N+} \sim \sqrt{\sigma}$ and $V_{\dc,\g}^{\D+} \sim \sigma^{-3/2}$. In the narrowband limit, $\sigma \to \infty$, Eqs.~(\ref{VdcGaussNNorm}) and (\ref{VdcGaussDNorm}) reduce to
\begin{subequations}\label{VdcGaussNorm}
\begin{align}
V_{\dc,\g}^{\N+}  &= \frac{1}{\omega_c R C_{0} \sqrt{\sigma}}\sqrt{\frac{2E_v}{\sqrt{\pi}}}, \label{VdcN}\\
V_{\dc,\g}^{\D+}  &= \frac{\omega_c L}{\sqrt{\sigma} R} \sqrt{\frac{2E_v}{\sqrt{\pi}} }. \label{VdcD}
\end{align}
\end{subequations}
Both $V_{\dc,\g}^{\N+}$ and $V_{\dc,\g}^{\D+}$ decrease as $ 1/\sqrt{\sigma}$ [Fig.~\ref{figS3}(a)--(c)]. A similar behavior is observed for the negative branch, $V_{\dc,\g}^{-}$, in both broadband and narrowband regimes [Fig.~\ref{figS3}(d)--(f)]. Thus, increasing the bandwidth of the pulse requires a higher minimum biasing level, which diverges as $\sigma \to 0$, whereas moving toward the narrowband regime reduces the required biasing, which approaches zero as $\sigma \to \infty$. In the narrowband case, the system approaches the sinusoidal regime through increasingly broader pulses, and, when $\omega_0=\omega_c$, the standard resonance matching condition, $C(t) \to C_0$, is expected to be recovered. This behavior in the narrowband limit will be discussed in the next section.
\subsection{Behavior of $C(t)$ in the Narrowband Limit}
We consider the narrowband limit $\sigma \to \infty$, where the excitation consists of increasingly broad pulses approaching the sinusoudal regime. In this limit, the expression of $C(t)$ in Eq.~(\ref{CtGauss}) reduces to
\begin{equation}\label{CtGaussAsym}
C\left(t\right) = C_0 \frac{V_{\dc}-A\frac{1}{\omega_c R C_0} \cos\!\left(\omega_c (t-\mu)\right)}{V_{\dc}-A\frac{\omega_c L}{R}\cos\!\left(\omega_c (t-\mu)\right)},
\end{equation}
and, combining $V_{\dc,\g}^{+} >\max\!\left\{V_{\dc,\g}^{\N+},\, V_{\dc,\g}^{\D+}\right\}$ with Eqs.~(\ref{VdcNPGaussAsymNar}) and (\ref{VdcDPGauss}) gives
\begin{equation}\label{}
V_{\dc,\g}^{+} >A\omega_c \frac{L}{R}\max\!\left\{\frac{\omega_0^2}{\omega_c^2},\, 1\right\}.
\end{equation}
To facilitate the analysis, the previous inequality can be parameterized as $V_{\dc,g}^{+} =b A\omega_c \frac{L}{R}\max\!\left\{\frac{\omega_0^2}{\omega_c^2},\, 1\right\},\; b>1$. Substituting this parameterized expression for $V_{\dc,\g}^{+}$ in Eq.~(\ref{CtGaussAsym}), we obtain $C(t)$ as
\begin{subequations}\label{CtGaussPW}
\begin{align}
C(t) &= C_0 \frac{b-\cos\!\left(\omega_c (t-\mu)\right)}{b-\frac{\omega_c^2}{\omega_0^2}\cos\!\left(\omega_c (t-\mu)\right)}, && \omega_0 \ge \omega_{c}, \label{}\\
C(t)  &= C_0 \frac{b-\frac{\omega_0^2}{\omega_c^2} \cos\!\left(\omega_c (t-\mu)\right)}{b-\cos\!\left(\omega_c (t-\mu)\right)}, && \omega_0 \le \omega_{c}. \label{}
\end{align}
\end{subequations}
From these expressions, we observe that when $\omega_0=\omega_c$, $C(t)=C_0$ and the standard resonance matching condition is recovered. When $\omega_0 \neq \omega_c$, however, even in the narrowband limit the standard resonance matching condition is not satisfied, achieving matching requires $C(t)$ to vary in time and to be appropriately biased so that it remains finite and strictly positive.

\section{Derivation of Lower Bounds on the Minimum Required Biasing Voltage}
In this section, we extend the analysis by deriving a general lower bound on the minimum required biasing voltage that is not tied to a specific temporal waveform. The resulting bound ensures distortionless impedance matching and depends only on the pulse bandwidth and energy, remaining independent of its temporal profile. To this end, we assume that the incoming pulse, $v^+\left(t\right)$, is real-valued, passband, and of finite energy $E_v=\|v^+\left(t\right)\|_2^2=\int_{-\infty}^{\infty}|v^+\left(t\right)|^2\,dt<\infty$ and with spectral support in $[-\omega_2,-\omega_1]\cup[\omega_1,\omega_2]$, $0<\omega_1<\omega_2$. As discussed above, ensuring a physically realizable capacitance, $C(t)>0$ for all $t$, imposes a constraint on the biasing voltage, which must satisfy $V_{\dc}^{+} >\max\!\left\{V_{\dc}^{\N+},\, V_{\dc}^{\D+}\right\}$, where $V_{\dc}^{\N+}$ and $V_{\dc}^{\D+}$ are defined in Eqs.~(\ref{VdcPN}) and (\ref{VdcPD}), respectively. Therefore, deriving upper bounds on $V_{\dc}^{\N+}$ and $V_{\dc}^{\D+}$ yields a lower bound on $V_{\dc}^{+}$. Since the derived bound is symmetric with respect to the sign of $V_{\dc}$, as discussed in the main text, we omit the superscript “$+$” in the following for notational simplicity. We begin with $V_{\dc}^{\D}$ [Eq. (\ref{VdcPD})], which can be equivalently expressed as
\begin{equation}\label{VdcD1}
V_{\dc}^{\D}=\frac{L}{R}\left\| \frac{d v^+(t)}{dt}\right\|_\infty
\end{equation}
where $\|\cdot\|_\infty$ denotes the supremum norm. Applying Bernstein’s inequality \cite{lapidoth2009SM,pinsky2002SM} to the right-hand side of this expression, the derivative of the pulse can be upper-bounded in terms of its maximum frequency, yielding
\begin{equation}\label{VdcD2}
\left\| \frac{d v^+(t)}{dt} \right\|_\infty \le \omega_2 \|v^+\left(t\right)\|_\infty.
\end{equation}
By invoking the Cauchy-Schwarz inequality \cite{lapidoth2009SM}, the supremum norm of the pulse can be related to its energy,
\begin{equation}\label{VdcD3}
\|v^+\left(t\right)\|_\infty \le \sqrt{\left(\omega_2-\omega_1\right)} \sqrt{\frac{E_v}{\pi}}.
\end{equation}
Combining Eqs.~(\ref{VdcD2}) and (\ref{VdcD3}) with Eq.~(\ref{VdcD1}), we obtain the upper bound on $V_{\dc}^{\D}$:
\begin{equation}\label{VdcD}
V_{\dc}^{\D}  \le \omega_2 \frac{L}{R}\sqrt{\omega_2-\omega_1} \sqrt{\frac{E_v}{\pi}}.
\end{equation}

We now derive an upper bound on $V_{dc}^{\N}$ [Eq.~(\ref{VdcPN})], which can be conveniently rewritten as
\begin{equation}\label{VdcN2}
V_{\dc}^{\N}  =  -\frac{1}{R C_{0}}\operatorname*{inf}_{t\in \mathbb{R}}  I(t),
\end{equation}
where $I(t)=\int_{-\infty}^{t} v^{+}(\nu)\, d\nu$. The right-hand side of Eq.~(\ref{VdcN2}) can be upper-bounded in terms of the supremum norm as
\begin{equation}\label{VdcN3}
-\frac{1}{R C_{0}}\operatorname*{inf}_{t\in \mathbb{R}}  I(t) \le \frac{1}{R C_{0}} \|I\left(t\right)\|_\infty.
\end{equation}
By bounding $I(t)$ via Fourier inversion and the Cauchy-Schwarz inequality, we obtain the following upper bound on $V_{\dc}^{\N}$:
\begin{equation}\label{VdcN}
V_{\dc}^{\N}  \le \frac{1}{C_0 R} \frac{\sqrt{\omega_2-\omega_1}}{\omega_1} \sqrt{\frac{E_v}{\pi}}.
\end{equation}
Combining Eqs.~(\ref{VdcD}) and (\ref{VdcN}) with the condition $V_{\dc}>\max\!\left\{V_{\dc}^{\N},\, V_{\dc}^{\D}\right\}$, we obtain the following lower bound on the required biasing voltage on $V_{\dc}$:
\begin{subequations}\label{VdcB}
\begin{align}
|V_{\dc}|  &> \frac{\sqrt{\omega_2-\omega_1}}{\omega_1 C_0 R} \sqrt{\frac{E_v}{\pi}},  && \omega_0 \ge \sqrt{\omega_2\omega_1}, \label{}\\
|V_{\dc}|  &> \frac{\omega_2 L \sqrt{\omega_2-\omega_1}}{R} \sqrt{\frac{E_v}{\pi}}, && \omega_0< \sqrt{\omega_2\omega_1}.  \label{}
\end{align}
\end{subequations}

\section{Derivation of the energy balance including modulation work}
The derivation proceeds from the instantaneous power balance of the time-varying matching network, which is subsequently integrated over the interaction interval to obtain the global energy relation. Let $t_0$ denote the arrival time of the incident pulse at the load ($z=0$). The instantaneous power in the transmission line at the input of the circuit is given by
\begin{equation}\label{p0}
p_{0}(t)=V_{\dc}i^+(t)+p_v(t)
\end{equation}
where $p_v(t)=v^+(t)i^+(t)$ denotes the instantaneous power of the incident current with $i^+(t)=v^+(t)/R$ denoting its current. The total instantaneous power in the circuit is
\begin{equation}\label{pt}
p(t)=\frac{d}{dt}\left(U_C(t)+U_L(t)\right)+p_R(t),
\end{equation}
where $U_C(t)=\frac{1}{2}C(t)v_C^2(t)$ is the capacitor stored energy, $U_L(t)=\frac{1}{2}Li_R^2(t)$ is the inductor stored energy, and $p_R(t)=v_R(t)i_R(t)$ is the instantaneous power dissipated in the resistor. In the reflectionless case, the instantaneous modulation power is therefore defined as
\begin{equation}\label{pmod}
p_\mathrm{mod}(t)=p(t)-p_{0}(t).
\end{equation}
With $C(t)$ given by Eq.~(1) in the main text, we have $v_C(t) = V_{\dc} - \frac{L}{R}\frac{dv^+(t)}{dt}$, $i_R(t)=i^+(t)$, and $p_R(t)=p_v(t)$. Consequently, by using the capacitor relation $q=Cv_c$, where $q$ denotes the capacitor charge, and the charge-conservation law $i^+=dq/dt$, the instantaneous modulation power reduces to
\begin{equation}\label{pmod_ref}
p_\mathrm{mod}(t)=-\frac{1}{2}V_{\dc} \frac{dq}{dt}+\frac{1}{2}L\left( \frac{dq}{dt}\frac{d^2q}{dt^2}-q\frac{d^3q}{dt^3}\right)
\end{equation}
The energy exchange with the system by the modulation source from $t_0$ to $t$ is $\Delta W_\mathrm{mod}(t)=\int_{t_0}^{t} p_\mathrm{mod}(\nu)\, d\nu$. Combining this expression with Eq.~(\ref{pmod_ref}) and noticing that $q$ is differentiable and $q(t_0)=q_0$, where $q_0$ denotes the capacitor charge for $t<t_0$, the modulation energy is obtained as
\begin{equation}\label{DeltaWmod}
\Delta W_{mod}(t)=-\frac{1}{2}V_{dc} \left(q(t)-q_0\right)+\frac{1}{2}L\left[\left( \frac{dq(t)}{dt}\right)^2-q(t)\frac{d^2q(t)}{dt^2}\right].
\end{equation}
Since we consider incoming pulses ($v^+\left(t,z\right)$) with finite temporal support, after the transient ($t \gg t_0$) the charge $q(t)$ returns to a constant value, say $q_\infty$, and becomes time-independent. In this limit, the previous expression corresponds to the total modulation energy exchanged with the system and reduces to
\begin{equation}\label{Wmod}
W_{mod}=-\frac{1}{2}V_{dc} \left(q_\infty-q_0\right).
\end{equation}

\section{Synchronization-Error Formulation and Analysis}
As demonstrated in the preceding sections, aperiodic time modulation enables broadband, distortion-free impedance matching well beyond the conventional Bode–Fano bound. This performance, however, relies on synchronization between the incident waveform and the prescribed capacitance profile, representing an important practical constraint of the proposed approach. In practical implementations, synchronization may be affected by multiple nonidealities, resulting in temporal misalignment and a nonzero reflected wave. To quantify this sensitivity, we introduce a synchronization error $\tau$ by replacing the ideal capacitance $C(t)$ with $C_{\tau}(t)=C(t-\tau)$ where $\tau$ denotes time-shifted modulation. We first derive the governing equations for the reflected waveform. The resulting formulation is then evaluated numerically for a Gaussian-enveloped sinusoidal incident pulse to determine the reflected waveform, normalized reflected energy, and degradation in matching performance as a function of the synchronization error.

We define the charge perturbation induced by the synchronization error as
\begin{equation}
    \Delta q_{\tau}(t)
    =
    q_{\tau}(t)-q(t),
    \label{eq:Dqdef}
\end{equation}
where \(q_{\tau}(t)\) is the capacitor charge associated with the shifted
capacitance profile $C_{\tau}(t)$, whereas $q(t)$ is the
capacitor charge under perfectly synchronized, reflectionless operation. The reflected voltage associated with a synchronization error \(\tau\) is
denoted by \(v^{-}(t;\tau)\). For notational simplicity, its dependence on
\(\tau\) is suppressed in the following derivation, and we write
\(v^{-}(t)\equiv v^{-}(t;\tau)\). Let \(i_{\tau}(t)\) denote the current
flowing into the load when the capacitance profile is shifted by \(\tau\).
Since
$\frac{dq_{\tau}(t)}{dt}=i_{\tau}(t)=\frac{v^{+}(t)-v^{-}(t)}{R}$,
whereas, under perfectly synchronized and reflectionless operation,
$\frac{dq(t)}{dt}=i^{+}(t)=\frac{v^{+}(t)}{R}$,
differentiating Eq.~\eqref{eq:Dqdef} yields
\begin{equation}
    \frac{d\Delta q_{\tau}(t)}{dt}
    =
    -\frac{v^{-}(t)}{R},
    \label{eq:loadcurrent}
\end{equation}
which directly relates the charge perturbation to the reflected voltage. Applying the load KVL equation, Eq.~\eqref{KVL}, to the perfectly
synchronized and temporally shifted cases and subtracting the former
from the latter gives
\begin{equation}
    \frac{d v^{-}(t)}{dt}
    =
    -\frac{2R}{L}v^{-}(t)
    +\frac{R}{L}q(t)
    \left[
        \frac{1}{C_{\tau}(t)}
        -\frac{1}{C(t)}
    \right]
    +\frac{R}{L}
    \frac{\Delta q_{\tau}(t)}{C_{\tau}(t)},
    \label{eq:error}
\end{equation}
Eqs.~\eqref{eq:loadcurrent} and \eqref{eq:error} constitute a
two-state, first-order, linear time-varying model describing the
synchronization-error dynamics, with \(v^{-}(t)\) and
\(\Delta q_{\tau}(t)\) as the state variables.
Eq.~\eqref{eq:error} identifies two contributions to the reflected wave. The first is the direct forcing caused by the temporal displacement of the capacitance profile. The second is a memory term associated with the accumulated deviation of the capacitor charge from the perfectly synchronized solution. The DC bias enters through $q(t)$ [Eq.~\eqref{eq:q}] and can therefore affect timing sensitivity.

When the shifted capacitance profile coincides with the ideal one, \(C_{\tau}(t)=C(t)\), the direct forcing term in Eq.~\eqref{eq:error} vanishes. As defined in the previous section, \(t_0\) denotes an instant preceding the arrival of the incident pulse. For the synchronization-error analysis, \(t_0\) is further chosen to precede the time at which the ideal and shifted capacitance profiles begin to differ. Thus, the two cases share the same initial state, such that \(v^{-}(t_0)=0\) and \(\Delta q_{\tau}(t_0)=0\). The unique solution of the resulting homogeneous error model is therefore \(v^{-}(t)=0\) and \(\Delta q_{\tau}(t)=0\), corresponding to perfectly synchronized, reflectionless operation.

For sufficiently small synchronization errors, and assuming that $C(t)$ is sufficiently smooth and nonzero, the shifted capacitance can be Taylor-expanded around $\tau=0$, while the reflected voltage and charge perturbation are represented by first-order perturbation expansions in $\tau=0$:
\begin{align}
    \frac{1}{C_{\tau}(t)}
    &=
    \frac{1}{C(t)}
    +\tau\frac{\dot C(t)}{C^2(t)}
    +O(\tau^2),\\
    v^{-}(t;\tau)
    &=
    \tau v_1^{-}(t)+O(\tau^2),
    \label{eq:vminus_first_order}\\
    \Delta q_{\tau}(t)
    &=
    \tau\Delta q_1(t)+O(\tau^2).
\end{align}
where $v_1^{-}(t)=\left.\frac{\partial v^{-}(t;\tau)}{\partial\tau}\right|_{\tau=0}$ and $    \Delta q_1(t) = \left.\frac{\partial\Delta q_{\tau}(t)}{\partial\tau}\right|_{\tau=0}$ are the first-order timing sensitivities. Substitution into the synchronization-error model and collection of the terms proportional to $\tau$ gives
\begin{align}
    \frac{d\Delta q_1(t)}{dt}
    &=
    -\frac{v_1^{-}(t)}{R},\\
    \frac{dv_1^{-}(t)}{dt}
    &=
    -\frac{2R}{L}v_1^{-}(t)
    +\frac{R}{L}\frac{\Delta q_1(t)}{C(t)}
    +\frac{R}{L}q(t)\frac{\dot C(t)}{C^2(t)},
\end{align}
The first-order approximation shows that the reflected-voltage waveform is first order in the timing error. To quantify the corresponding degradation in matching performance, we define the normalized reflected energy as
\begin{equation}
    \mathcal{R}_{E}(\tau)
    =
    \frac{\displaystyle\int_{t_0}^{t_f}
    \left|v^{-}(t;\tau)\right|^2\,dt}
    {\displaystyle\int_{t_0}^{t_f}
    \left|v^{+}(t)\right|^2\,dt}.
\end{equation}
Here, $t_f$ is selected sufficiently late that $v^{+}(t)$ and $v^{-}(t;\tau)$ are negligible for $t\geq t_f$ over the entire range of timing errors considered. Using Eq.~\eqref{eq:vminus_first_order}, the normalized reflected energy becomes
\begin{equation}
\mathcal{R}_{E}(\tau) = \mathcal{S}_{E}\tau^2+O(\tau^3),
\qquad
\mathcal{S}_{E}=\frac{\displaystyle\int_{t_0}^{t_f}\left|v_1^{-}(t)\right|^2\,dt}{\displaystyle\int_{t_0}^{t_f}\left|v^{+}(t)\right|^2\,dt}.
\end{equation}
Consequently, $\mathcal{R}_{E}(0)=0$ and $\left.d\mathcal{R}_{E}/d\tau\right|_{\tau=0}=0$, showing that the normalized reflected energy is stationary at perfect synchronization. Therefore, sufficiently small synchronization errors do not produce a first-order degradation in the energy-based matching performance; instead, the normalized reflected energy increases quadratically with the timing error. Perfect synchronization is thus first-order insensitive, in an energy sense, to small timing perturbations, while the coefficient $\mathcal{S}_{E}$ determines the rate at which the matching performance degrades as the timing error increases.

\begin{figure}[ht]
  \centering
  \includegraphics[width=0.8\linewidth]{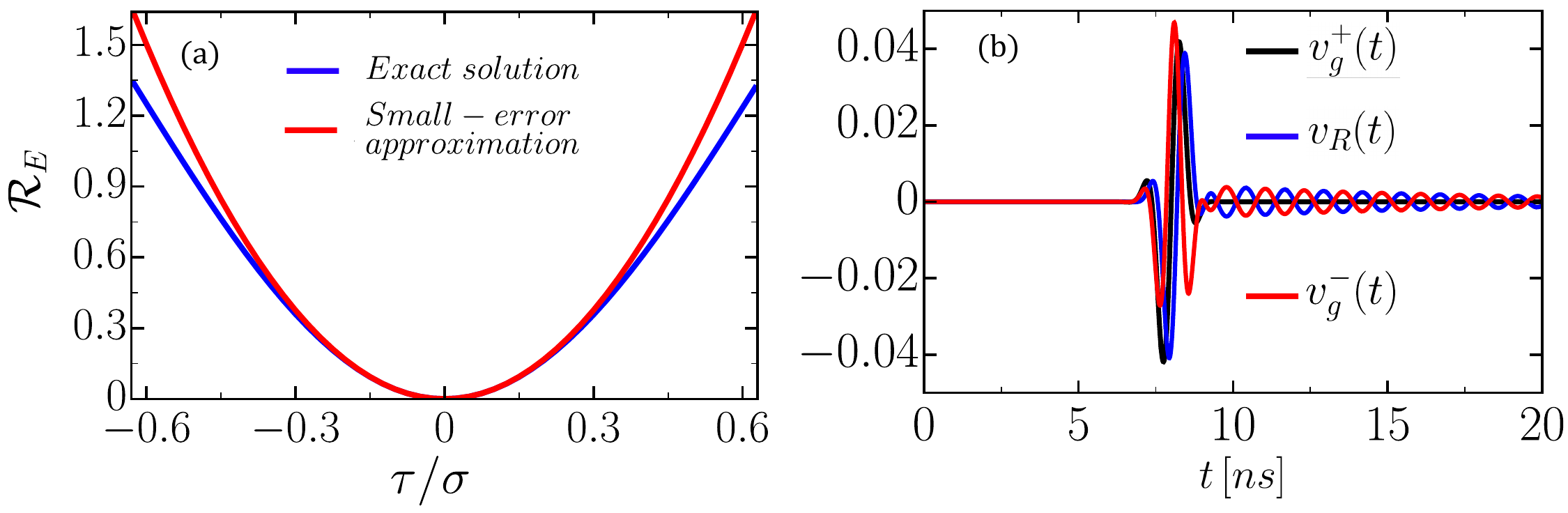}
  \caption{Synchronization-error response for a Gaussian-enveloped sinusoidal pulse. (a) Normalized reflected energy $\mathcal{S}_{E}$ versus $\tau/\sigma$ from the exact model and first-order approximation. (b) Incident, reflected, and load-resistor voltages for $\tau/\sigma=0.5$. The results were obtained for $E_v=1$ [pJ], $\sigma=0.4$ [ns], $L=10$ [nH], $R=1~[\Omega]$, $\omega_c=5$ rad/ns, and $\omega_0=1/\sqrt{C_0 L}$.}
  \label{figSyncError}
\end{figure}

We next evaluate the synchronization sensitivity of the proposed matching scheme for the Gaussian-enveloped sinusoidal incident pulse [Eq.~\eqref{VGauss}] by comparing the exact response with the quadratic small-error approximation. As shown in Fig.~\ref{figSyncError}(a), the first-order model accurately reproduces the exact normalized reflected energy for sufficiently small values of $\sigma/\tau$, while the discrepancy increases as the timing error becomes larger. Note that the normalized reflected energy exceeds unity for sufficiently large timing errors. This behavior is possible because the system is time modulated, allowing the modulation and bias mechanisms to exchange energy with the circuit and transfer part of that energy to the outgoing wave. To complement the energy-based analysis, Fig.~\ref{figSyncError}(b) shows the temporal profiles of the incident voltage $v_g^{+}(t)$, the reflected voltage $v_g^{-}(t)$, and the voltage across the load resistor $v_R(t)$ for a representative synchronization error. The timing mismatch produces a nonzero reflected waveform and a corresponding deviation of the resistor voltage from the incident pulse. The comparison provides a direct view of how the reflected waveform modifies the voltage delivered to the load resistor. Both the reflected voltage and the resistor voltage exhibit post-pulse ringing associated with the residual transient dynamics of the circuit.

% ==========================
% References (if needed)
% ==========================
% If your main paper uses BibTeX, you can reuse the same .bib file here.
% Choose one of the following approaches.

% --- Approach A: BibTeX ---
%\bibliographystyle{apsrev4-2}
%\bibliography{mybib_sup} % <- your .bib filename without extension

% --- Approach B: Manual bibliography (comment out Approach A if using this) ---
% \begin{thebibliography}{99}
% \bibitem{key1} Author, \textit{Title}, Journal \textbf{Volume}, Page (Year).
% \end{thebibliography}

%\end{document}